\documentclass[12pt,a4paper]{article}
\usepackage{amsmath,amssymb,amsfonts}
\usepackage{graphicx}
\usepackage{booktabs,colortbl}

\newcommand{\be}{\begin{equation}} 
\newcommand{\ee}{\end{equation}} 
\newcommand{\bea}{\begin{eqnarray}}  
\newcommand{\eea}{\end{eqnarray}}
\newcommand{\bs}{\begin{split}} 
\newcommand{\es}{\end{split}}

\newcommand{\units}[1]{~\mathrm{#1}}

\newcommand{\scriptmath}[1]{{\scriptsize{\mbox{$#1$}}}}

\newcommand{\hc}{\mathrm{h.c.}} 
\newcommand{\mt}[1]{\mathrm{#1}}

\begin{document}

%----------------------------------- TITLE AND AUTHORS -----------------------------------------%

%Preprint numbers
\begin{flushright}
%\today
\end{flushright}
\vspace*{5mm}
\begin{center}

\renewcommand{\thefootnote}{\fnsymbol{footnote}}

{\Large {\bf Singlet deflected anomaly/gauge mediation
}} \\
\vspace*{0.75cm}
{\bf J.\ de Blas}\footnote{E-mail: jdeblasm@nd.edu}
and
{\bf A.\ Delgado}\footnote{E-mail: antonio.delgado@nd.edu}

\vspace{0.5cm}

Department of Physics, University of Notre Dame,\\
Notre Dame, IN 46556, USA

\end{center}
\vspace{.5cm}

%--------------------------------------------- ABSTRACT ---------------------------------------------%
\begin{abstract}
 
\noindent We study an extension of the standard anomaly/gauge mediation scenario where the messenger fields have direct interactions with an extra gauge singlet. This realizes a phenomenologically viable NMSSM-like scenario free of the $\mu$-$b_\mu$ problem. Current cosmological constraints imply a small size for the anomaly-mediation contributions, unless some source of $R$-parity violation is permitted. In the latter case
the allowed regions in the parameter space can be substantially larger than in the corresponding gauge-mediation scenario.

\end{abstract}

\renewcommand{\thefootnote}{\arabic{footnote}}
\setcounter{footnote}{0}

%-------------------------------- DOCUMENT: INTRODUCTION ---------------------------------%

\section{Introduction}
\label{section_Intro}

Theories with supersymmetry (SUSY) are one of the most appealing candidates for physics beyond the Standard Model (SM), providing an elegant  solution to the hierarchy problem. Should SUSY be realized in nature, we expect to eventually find its tracks at the large hadron collider (LHC). In such a case new questions arise. In particular, we need to understand why SUSY is not preserved at low energies. As a bonus, one might expect that the knowledge of the SUSY breaking mechanism may also shed light over some of the known problems of supersymmetric extensions of the SM, like the $\mu$ or the flavor problems. For instance, explaining why $\mu$ --the Higgsino mass parameter-- has to be of the right (electroweak) size implies that this must not be present in the exact supersymmetric limit, for it would be naturally of the order of the Planck or another large fundamental scale. Thus, $\mu$ must be generated upon SUSY breaking and its origin may be intimately related to the mechanism responsible of the breakdown.

A quick review of the standard SUSY breaking mechanisms would include, for instance, gravity mediation \cite{Chamseddine:1982jx}, gauge mediation \cite{Dine:1981gu} or anomaly mediation \cite{Randall:1998uk,Giudice:1998xp} of SUSY breaking. In gravity mediation SUSY breaking is communicated to the visible sector at tree level, by means of non-renormalizable interactions suppressed by the Planck scale, $M_P$. In this kind of scenarios all soft parameters are generated at the same order,
\be
M_{\lambda_a}\sim m_{\varphi_i} \sim a_y\sim \frac{F}{M_P},
\label{GravMSB}
\ee
with $F$ the F-term vacuum expectation value (vev) of the hidden-sector chiral superfields, $X$,  responsible of SUSY breaking.

Gravity mediated models address in an elegant way the $\mu$ problem \cite{Giudice:1988yz}. This is generated at the same order than the other soft masses and,  in particular, $b_\mu \sim \mu^2$, with $b_\mu$ the Higgs soft bilinear term in the scalar potential. This follows from the following interactions in the K\"ahler potential,
\be
K\supset \frac{a}{M_P}X^\dagger H_u\cdot H_d +\frac{b}{M_P^2}X^\dagger X H_u\cdot H_d +\hc,
\ee
and from the fact that in gravity mediation both $a$ and $b$ can be naturally of order one. On the other hand, in general gravity mediation models come with the flavor problem. Indeed, the most general K\"ahler potential can include flavor-violating non-renormalizable interactions such as, for instance, $\frac{c_{ij}}{M_P^2}X^\dagger X \varphi^\dagger_i \varphi_j$, which yield non-universal contributions to soft parameters, and in particular to the sfermion soft masses squared. 

There is no such flavor problem in models with pure gauge mediation of SUSY breaking (GMSB). In this case SUSY breaking is communicated to the visible sector by the gauge interactions of a set of messenger fields. SUSY breaking in the hidden sector is parametrized by a spurion field with vev $\left<X\right>=M+F\theta^2$. This provides non-supersymmetric masses to the messenger fields, whose  r{\^o}le is usually played by a given number of  pairs $(\Phi,\bar{\Phi})$ in the $(5,\bar{5})$ of $SU(5)$. Such choice guarantees that gauge unification is preserved. Contributions to the soft terms for the visible sector are induced at the loop level as follows:
\be
M_{\lambda_a} \sim \frac{g_a^2}{16\pi^2}\frac{F}{M},~~~~~~~~
m_{\varphi_i}^2\sim\sum_a\frac{g_a^4}{256\pi^4}\frac{F^2}{M^2}\sim M_\lambda^2,~~~~~~~~
a_y\approx 0.
\label{GMSB}
\ee
In this case, however, the $\mu$ problem translates into a hierarchy problem. In this class of models it is not difficult to obtain $\mu$ of the right order. The problem comes from the fact that the same interactions generating $\mu$ also generate $b_\mu$ at the same order in the loop expansion, in a way that
\be
\frac{b_\mu}{\mu}\sim \frac{F}{M}.
\ee
This is parametrically too large since, in order to push sparticle masses above current bounds, $F/M$ must be $\sim 10$ - $100\units{TeV}$. This would require an unnatural fine tuning to explain electroweak symmetry breaking (EWSB). 

Anomaly mediation is also free of the flavor problem and its contributions to soft terms are present in any supergravity (SUGRA) model. They have their origin in the superconformal anomaly and can be described by the vev of a chiral superfield $\phi$, acting as a compensator of the super Weyl transformations \cite{Cremmer:1978hn}. Setting $\left<\phi\right>=1+F_\phi\theta^2$, the gravitino mass, $m_{3/2}\sim F_\phi$, is the only new parameter controlling the contributions to soft parameters. These are given by
\be
M_{\lambda_a}=\frac{\beta_{g_a}}{g_a}m_{3/2},~~~~~~~~
m^2_{\varphi_i}=-\frac{1}{4}\left(\frac{\partial \gamma_{\varphi_i}}{\partial g_a}\beta_{g_a}+\frac{\partial \gamma_{\varphi_i}}{\partial y}\beta_{y}\right)m_{3/2}^2,~~~~~~~~
a_y=-\beta_y m_{3/2},
\label{AMSB}
\ee
with $\beta_i$, $\gamma_i$ the beta functions for the couplings and anomalous dimensions of the fields, respectively.
A nice feature of pure anomaly mediation of SUSY breaking (AMSB) is that the above equations hold at all scales. This results in an highly predictive scenario, since all soft masses can be computed in terms of the gravitino mass and the low-energy parameters of the theory. This high predictive power, however, turns against the standard AMSB since the contributions to soft masses squared for the first two slepton families turn out to be negative. This tachyonic slepton problem is a phenomenological disaster, as it would destabilize the $U(1)_\mt{em}$ invariant vacuum. The $\mu$-$b_\mu$ problem is also present in AMSB. A pure $\mu$ term is strictly forbidden in the exact SUGRA theory since it would explicitly break the Weyl invariance. Introducing it in the superpotential generates also a hierarchy $b_\mu/\mu=m_{3/2}$, which is again too large. 

As can be seen, none of the different mechanisms of supersymmetry breaking are completely satisfactory by themselves. Of course, it may also happen that not only one but several different mechanisms contribute at the same time. In particular, AMSB must be present as long as SUGRA is realized as an effective theory at some point. In this regard, several different scenarios have been proposed in order to solve the tachyonic slepton problem of anomaly mediation (see, for instance, \cite{Pomarol:1999ie,AnomalyPlus}). One of the first attempts was to combine anomaly with gauge mediation  of SUSY breaking (AGMSB) \cite{Pomarol:1999ie}. Indeed, since contributions to soft masses squared arise at the same order in the loop expansion it is possible to balance the AMSB tachyonic slepton masses with the positive GMSB contributions.

This combined scenario of SUSY breaking is free from tachyonic and flavor problems but still suffers from the $\mu$-$b_\mu$ problem. As in each separate model, this can be addressed within the Next-to Minimal Supersymmetric Standard Model (NMSSM).  Because of the existence of a discreet $\mathbb{Z}_3$  symmetry, broken only at the weak scale, in the NMSSM the $\mu$ term is only generated effectively through the vev of a new gauge singlet $S$ with superpotential interactions
\be
W\supset \lambda S H_u\cdot H_d -\frac{\kappa}{3}S^3~\longrightarrow~\mu^\mt{Eff}=\lambda \left<S\right>.
\ee
The effective $b_\mu$ term is given by $b_\mu^\mt{Eff}=a_\lambda \left<S\right>+\lambda \kappa \left<S\right>^2$, with $a_\lambda$ the Higgs-singlet trilinear soft coupling in the soft scalar potential. Thus, $b_\mu/\mu\sim a_\lambda/\lambda+\kappa\left<S\right>\sim m_{3/2}/16\pi^2+\kappa\left<S\right>$, solving the problem. The question is whether we can achieve EWSB in this scenario or not, and if the resulting spectrum is phenomenologically viable. EWSB in general requires a large vev for the singlet. This needs of a negative $m_S^2$ and/or large $a$ terms for the singlet superpotential interactions, $a_\lambda$ and $a_\kappa$. It is well known that these are not possible in the NMSSM with the standard GMSB \cite{deGouvea:1997cx} or AMSB \cite{Kitano:2004zd}. In the gauge-mediation side this is so because $S$ has no gauge quantum numbers and $a$ terms are typically very suppressed. On the other hand, AMSB always predicts a positive value for $m_S^2$. In this case, $a_\lambda$ and $a_\kappa$ can be sizable, but only if $\lambda$ and $\kappa$ are also large. Such large values, however, would induce a larger $m_S^2$, suppressing the singlet vev. Following \cite{Giudice:1997ni,Delgado:2007rz,deBlas:2011hs} one can add extra superpotential interactions between the singlet fields and $n=2$ (or in general an even number) pairs of messengers, to help in attaining EWSB:
\be
W_\mt{mess}\supset \xi S \bar{\Phi}_1 \Phi_2.
\ee
As discussed in \cite{Delgado:2007rz}, this kind of interactions allows to obtain a realistic EWSB in the gauge mediated NMSSM. The absence in \cite{Delgado:2007rz} of direct contributions to the trilinear soft term associated to the top Yukawa coupling, however, results naturally in relatively small values of the stop mixing. This forces to push the messenger and the effective SUSY breaking scales ($M$ and $F/M$, respectively) up, in order to lift the lightest CP-even Higgs mass, $m_{H^1}$, above the LEP 2 bound. Otherwise, the model has an extra fine tuning, living in tiny regions of the parameter space. Thus, that implementation disfavors a relatively light spectrum. In this paper we address the problem for the AGMSB scenario. Since AMSB provides sizable (one-loop) contributions to $a$ terms, one would expect that singlet deflection of this joint scenario should naturally allow for a lighter spectrum.\footnote{This is a different approach to the one followed in \cite{deBlas:2011hs}, where the extra contributions to $a_t$ are provided within a gauge mediation like scenario. Compared to that paper, the model  described here is naturally more predictive. Indeed, although only one extra parameter was responsible of generating a sizable $a_t$ in \cite{deBlas:2011hs}, many extra new interactions were allowed by the symmetries and, in fact, were generated in renormalization.} As we will show, however, extra phenomenological constraints not applying in the gauge-mediation case have to be taken into account here. These impose significant restrictions on the size of the anomaly mediation contributions to soft terms.

In the next section we quickly review the theoretical aspects of the model. In section \ref{section_PhResults} we present our phenomenological results. These are obtained from a scan over the parameter space looking for the regions where EWSB is possible. We compare with the gauge-mediation scenario and explain the difficulties of the combined model. We discuss the resulting spectrum in turn. Finally we summarize in the conclusions.

%-------------------------------- DOCUMENT: SECTIONS -----------------------------------------%

\section{The model}
\label{sec:Model}

The superpotential of the model can be split in two pieces, $W=W_\mt{NMSSM}+W_\mt{mess}$. The NMSSM part, $W_\mt{NMSSM}$, contains the standard superpotential interactions:
\be
W_\mt{NMSSM}=-y_t~u_3^c \left(H_u \cdot q_3\right) +y_b~d_3^c \left(H_d \cdot q_3\right)+y_\tau~e_3^c \left(H_d \cdot l_3\right) +\lambda~S \left(H_u\cdot H_d\right)-\frac{\kappa}{3}~S^3.
\label{NMSSMW}
\ee
For the messenger sector we consider $n=2$ pairs of fields $(\Phi_i,\bar{\Phi}_i)$ in the $(5,\bar{5})$ of $SU(5)$. We assume they only couple to the (non-dynamical) spurion field $X$ and to the NMSSM gauge singlet $S$,
\be
\begin{split}
W_\mt{mess}&=X\sum_{i=1}^{n=2}\left(\kappa_i^D \bar{\Phi}^D_i\Phi^D_i+\kappa_i^T \bar{\Phi}^T_i\Phi^T_i\right)+S\left(\xi_D \bar{\Phi}^D_1\Phi^D_2+\xi_T \bar{\Phi}^T_1\Phi^T_2\right).
\end{split}
\label{MessW}
\ee
In Eq.~\!(\ref{MessW}) $\Phi_i^{D,T}$ and $\bar{\Phi}_i^{D,T}$ denote the corresponding $SU(2)_L$ doublet and $SU(3)_c$ triplet components of the messenger fields, respectively.
 
As shown in \cite{Delgado:2007rz} the above superpotential can by explained by extending the discrete $\mathbb{Z}_3$ symmetry of the NMSSM with $\mathbb{Z}_3[\Phi_1]=\mathbb{Z}_3[\bar{\Phi}_2]=-1/3$, $\mathbb{Z}_3[\Phi_2]=\mathbb{Z}_3[\bar{\Phi}_1]=1/3~(=\mathbb{Z}_3[S]=\mathbb{Z}_3[H_u]=\mathbb{Z}_3[H_d])$, $\mathbb{Z}_3[X]=0$. On the other hand, the two messengers are required in order to avoid kinetic mixing between $X$ and $S$, which could destabilize the weak scale. Note that there are other interactions that are still allowed by the symmetries of the problem (see \cite{deBlas:2011hs} for a description of such terms) but, compared to \cite{deBlas:2011hs}, the form of $W_\mt{mess}$ is preserved by the non-renormalization theorem.

SUSY breaking is parametrized by the the vevs of both, the conformal compensator field $\left<\phi\right>=1+m_{3/2} \theta^2$ and the spurion $\left<X\right>=M+F\theta^2$, which give rise to the anomaly and gauge mediation contributions, respectively. As usual in AMSB, potentially dangerous sources of flavor violation coming from contact terms of the form $\frac{c_{ij}}{M_P^2}X^\dagger X \varphi^\dagger_i \varphi_j$ can be suppressed assuming sequestering in an extra dimensional scenario. In that case, the locality in the extra dimensions can result in an exponential suppression of the gravity mediation contributions \cite{Randall:1998uk,Luty:1999cz}. 

Above the messenger mass scale all the soft parameters adjust to the AMSB renormalization group (RG) trajectory in Eq.~\!(\ref{AMSB}), with the $\beta_i$ and $\gamma_i$ functions computed in the theory including the messenger fields. Below the messenger mass threshold, after integrating the messenger fields out, the soft terms are corrected by the singlet-deflected gauge-mediation contributions and then deviate from the AMSB form. This fixes the values for the boundary conditions at $Q=M$, that then have to be evolved using the NMSSM RG equations. 

The gaugino masses at $Q=M$ are given by the AGMSB expression. At one loop,
\be
M_{\lambda_a}=\frac{g_a^2}{16\pi^2}\left(2\frac{F}{M}+b_{a}m_{3/2}\right),~~a=1,2,3,
\label{Ma}
\ee
where $b_a=(\frac{43}{5},3,-1)$ and we have used $n=2$ messengers. The expressions for the scalar soft masses squared are rather more involved. They receive the dominant ${\cal O}\left(F^2/M^2,m_{3/2}^2\right)$ contributions at two loops. The MSSM sfermions masses squared come only from AGMSB. For the third family:

\be
\begin{split}
m_{q_3}^2\!=&\frac{1}{256 \pi^4}\!\!\left[\!\left(\frac{g_1^4}{15}+3g_2^4+\frac{16}{3}g_3^4\right)\!\frac{F^2}{M^2}\!+\!\left(\!y_t^2\left(\!-\frac{13}{15}g_1^2-3g_2^2-\frac{16}{3}g_3^2+6y_t^2+y_b^2+\lambda^2\!\right)\!+\right.\right.\\
+&\left.\left.y_b^2\left(\!-\frac{7}{15}g_1^2-3g_2^2-\frac{16}{3}g_3^2+y_t^2+6y_b^2+y_\tau^2+\lambda^2\!\right)\!-\frac{43}{150}g_1^4-\frac{9}{2}g_2^4+\frac{8}{3}g_3^4\right)\!m_{3/2}^2\right],\\
\\[-0.25cm]
m_{u^c_3}^2\!=&\frac{1}{256 \pi^4}\!\!\left[\!\left(\frac{16}{15}g_1^4+\frac{16}{3}g_3^4\right)\!\frac{F^2}{M^2}\!+\!\left(\!2y_t^2\left(\!-\frac{13}{15}g_1^2-3g_2^2-\frac{16}{3}g_3^2+6y_t^2+y_b^2+\lambda^2\!\right)\!-\right.\right.\\
-&\left.\left.\frac{344}{75}g_1^4+\frac{8}{3}g_3^4\right)\!m_{3/2}^2\right],\\
\\[-0.25cm]
m_{d^c_3}^2\!=&\frac{1}{256 \pi^4}\!\!\left[\!\left(\frac{4}{15}g_1^4+\frac{16}{3}g_3^4\right)\!\frac{F^2}{M^2}\!+\!\left(\!2y_b^2\left(\!-\frac{7}{15}g_1^2-3g_2^2-\frac{16}{3}g_3^2+y_t^2+6y_b^2+y_\tau^2+\lambda^2\!\right)\!-\right.\right.\\
-&\left.\left.\frac{86}{75}g_1^4+\frac{8}{3}g_3^4\right)\!m_{3/2}^2\right],\\
\\[-0.25cm]
m_{l_3}^2\!=&\frac{1}{256 \pi^4}\!\!\left[\!\left(\frac{3}{5}g_1^4+3 g_2^4\right)\!\frac{F^2}{M^2}\!+\!\left(\!y_\tau^2\left(\!-\frac{9}{5}g_1^2-3g_2^2+3y_b^2+4y_\tau^2+\lambda^2\!\right)\!-\frac{129}{50}g_1^4-\frac{9}{2}g_2^4\right)\!m_{3/2}^2\right],\\
\\[-0.25cm]
m_{e^c_3}^2\!=&\frac{1}{256 \pi^4}\!\!\left[\frac{12}{5}g_1^4\frac{F^2}{M^2}\!+\!\left(\!2y_\tau^2\left(\!-\frac{9}{5}g_1^2-3g_2^2+3y_b^2+4y_\tau^2+\lambda^2\!\right)\!-\frac{258}{25}g_1^4\right)\!m_{3/2}^2\right].
\end{split}
\label{sf3mass}
\ee
The corresponding expressions for the first two generations of sfermions can be obtained by taking all the Yukawa couplings to zero in (\ref{sf3mass}). Neglecting RG effects, requiring that the light sfermion families are not tachyonic would translate into the following rough relation between the SUSY breaking scales: $F/M\gtrsim 2 m_{3/2}$. 

In addition to the AGMSB contributions, the soft masses for the Higgses and the scalar singlet $S$ receive extra terms from the singlet-messenger interactions in (\ref{MessW}):

\be
\begin{split}
m_{H_d}^2\!=&\frac{1}{256 \pi^4}\!\!\left[\left(\frac{3}{5}g_1^4+3g_2^4-2\lambda^2\left(\xi_D^2+\frac{3}{2}\xi_T^2\right)\right)\frac{F^2}{M^2}+\left(-\frac{129}{50}g_1^4-\frac{9}{2}g_2^4+\right.\right.\\
+&3y_b^2\left(-\frac{7}{15}g_1^2-3g_2^2-\frac{16}{3}g_3^2+y_t^2+6y_b^2+y_\tau^2+\lambda^2\right)\!+y_\tau^2\left(-\frac{9}{5}g_1^2-3g_2^2+3y_b^2+4y_\tau^2+\lambda^2\right)\!+\\
+&\left.\left.\lambda^2\left(-\frac{3}{5}g_1^2-3g_2^2+3y_t^2+3y_b^2+y_\tau^2+4\lambda^2+2\kappa^2+2\xi_D^2+3\xi_T^2\right)\right)m_{3/2}^2\right],\\
\\[-0.25cm]
m_{H_u}^2\!=&\frac{1}{256 \pi^4}\!\!\left[\left(\frac{3}{5}g_1^4+3g_2^4-2\lambda^2\left(\xi_D^2+\frac{3}{2}\xi_T^2\right)\right)\frac{F^2}{M^2}+\left(-\frac{129}{50}g_1^4-\frac{9}{2}g_2^4+\right.\right.\\
+&3y_t^2\left(-\frac{13}{15}g_1^2-3g_2^2-\frac{16}{3}g_3^2+6y_t^2+y_b^2+\lambda^2\right)+\\
+&\left.\left.\lambda^2\left(-\frac{3}{5}g_1^2-3g_2^2+3y_t^2+3y_b^2+y_\tau^2+4\lambda^2+2\kappa^2+2\xi_D^2+3\xi_T^2\right)\right)m_{3/2}^2\right],\\
\\[-0.25cm]
m_{S}^2\!=&\frac{1}{256 \pi^4}\!\!\left[\left(\!-\frac{6}{5}g_1^2\left(\xi_D^2+\frac{2}{3}\xi_T^2\right)-6g_2^2\xi_D^2-16g_3^2\xi_T^2-4\kappa^2\left(2\xi_D^2+3\xi_T^2\right)+8\xi_D^4+15\xi_T^4+\right.\right.\\
+&\left.12\xi_D^2\xi_T^2\right)\frac{F^2}{M^2}+\left(2\lambda^2\left(-\frac{3g_1^2}{5}-3g_2^2+3y_t^2+3y_b^2+y_\tau^2+4\lambda^2+2\kappa^2+2\xi_D^2+3\xi_T^2\right)+\right.\\
+&2\kappa^2\left(6\lambda^2+6\kappa^2+6\xi_D^2+9\xi_T^2 \right)+2\xi_D^2\left(-\frac{3g_1^2}{5}-3g_2^2+2\lambda^2+2\kappa^2+4\xi_D^2+3\xi_T^2\right)+\\
+&\left.\left.3\xi_T^2\left(-\frac{4g_1^2}{15}-\frac{16}{3}g_3^2+2\lambda^2+2\kappa^2+2\xi_D^2+5\xi_T^2\right)\right)m_{3/2}^2\right].
\end{split}
\label{HSmass}
\ee
Note that, because of the extra $\xi_{D,T}$ contributions in the $\beta$ functions, the AMSB prediction needs not be always positive and, in particular, can become negative for small values of $\lambda$, $\kappa$ and not too large values of $\xi_{D,T}$. Finally, the contributions to $a$ terms are dominated by anomaly mediation and the $\xi_{D,T}$ terms, which enter at the one-loop level:
\be
\begin{split}
a_t\!=&\!-\!\frac{y_t}{16\pi^2}\!\left(6 y_t^2+y_b^2+\lambda^2-\frac{13}{15}g_1^2-3g_2^2-\frac{16}{3}g_3^2\right)\!m_{3/2},\\
a_b\!=&\!-\!\frac{y_b}{16\pi^2}\!\left(y_t^2+6y_b^2+y_\tau^2+\lambda^2-\frac{7}{15}g_1^2-3g_2^2-\frac{16}{3}g_3^2\right)\!m_{3/2},\\
a_\tau\!=&\!-\!\frac{y_\tau}{16\pi^2}\!\left(3y_b^2+4y_\tau^2+\lambda^2-\frac{9}{5}g_1^2-3g_2^2\right)\!m_{3/2},\\
a_\lambda\!=&\!-\!\frac{\lambda}{16\pi^2}\!\left[\!\left(2\xi^2_D+3\xi^2_T\right)\!\frac{F}{M}\!+\!\left(\!3y_t^2+3y_b^2+y_\tau^2+4\lambda^2+2\kappa^2+2\xi_D^2+3\xi_T^2-\frac{3}{5}g_1^2-3g_2^2\right)\!m_{3/2}\!\right],\\
a_\kappa\!=&\!-\!3\frac{\kappa}{16\pi^2}\!\left[\!\left(2\xi_D^2+3\xi_T^2\right)\!\frac{F}{M}\!+\!\left(2\lambda^2+2\kappa^2+2\xi_D^2+3\xi_T^2\right)\!m_{3/2}\!\right].
\end{split}
\label{aterms}
\ee

As in \cite{Delgado:2007rz} the $\xi_{D,T}$ contributions, complemented in this case with those of anomaly mediation, can provide large enough contributions to $a_\lambda$ and $a_\kappa$, as required by EWSB. On the other hand, the pure anomaly-mediation contribution to $a_t$ could help in attaining in a natural way a sizable stop mixing (and then a large $m_{H^1}$) without relying on too large values of $F/M$ or a large RG running. Since the $a_t$ contribution is proportional to the $y_t$ beta function, this requires not only a large gravitino mass but also the messenger mass scale to be away from top Yukawa quasi-fixed point, for which $\beta_{y_t}\approx 0$.

\section{Phenomenological results}
\label{section_PhResults}

We have studied EWSB for this model, performing a scan over the whole parameter space. We follow the same scan procedure in \cite{deBlas:2011hs} and refer to that reference and \cite{Delgado:2007rz} for details. In short, we have to evolve all the input parameters (the SM and the new parameters) from the scales where they are defined up to the messenger mass scale. There we can use Eqs.~\!(\ref{Ma})-(\ref{aterms}) to compute the values of the boundary conditions. These must then be evolved down to the scale where EWSB takes place by using the corresponding NMSSM RG equations (see for instance \cite{Ellwanger:2009dp}). We choose to minimize the scalar potential at a scale given by the geometric average of the stop masses, $M_\mt{match}=\sqrt{m_{\tilde{t}_1}m_{\tilde{t}_2}}$. This minimizes the form of the leading ${\cal O}\left(y_t^4\right)$ corrections to the scalar potential. After using all the SM input parameters, as well as the constraints provided by requiring the existence of the electroweak vacuum, there are a total of six free parameters. These are the scales $M$, $F/M$, the gravitino mass $m_{3/2}$ and the superpotential couplings $\lambda$, $\xi_D$ and $\xi_T$. We can remove another parameter by assuming that $\xi_{D,T}$ unify at a common value $\xi_U$ at the grand unification scale\footnote{The corresponding $\beta$ functions for the region between the messenger and the GUT scales can be found in the Appendix A of \cite{Delgado:2007rz}.}, which is determined requiring $g_1(M_\mt{GUT})=g_2(M_\mt{GUT})$. After finding a point where the electroweak minimum is a global minimum of the scalar potential we compute the spectrum, and request that it pass the same experimental constraints described in \cite{deBlas:2011hs}. In this case, to implement the LEP constraints on the Higgs masses we use the package {\tt NMHDECAY} \cite{Ellwanger:2004xm}. For the predictions of the Higgs sector masses we consider the leading one and two-loop corrections, as in \cite{deBlas:2011hs}.

From the results of the scan we find\footnote{As in \cite{deBlas:2011hs} we are mainly interested in low scale SUSY breaking scenarios that could be eventually tested at the LHC. Therefore, we restrict the scan to values of $F/M \le 175 \units{TeV}$. This is also the upper bound considered for the gravitino mass. The messenger mass scale runs from $10^4$ to $10^{11} \units{TeV}$.}
\be
\frac{F}{M}\gtrsim  50\units{TeV},~~~~m_{3/2}\lesssim 150 \units{TeV},
\ee
for the gauge and anomaly mediation SUSY breaking scales, respectively. As we explain below, however, large values of $m_{3/2}$ are disfavored. In general, points with $m_{3/2}\gg 30\units{TeV}$ are difficult to obtain in the scan, and thus may not be considered as natural solutions.

\begin{figure}[t!]
% arXiv plots
\input{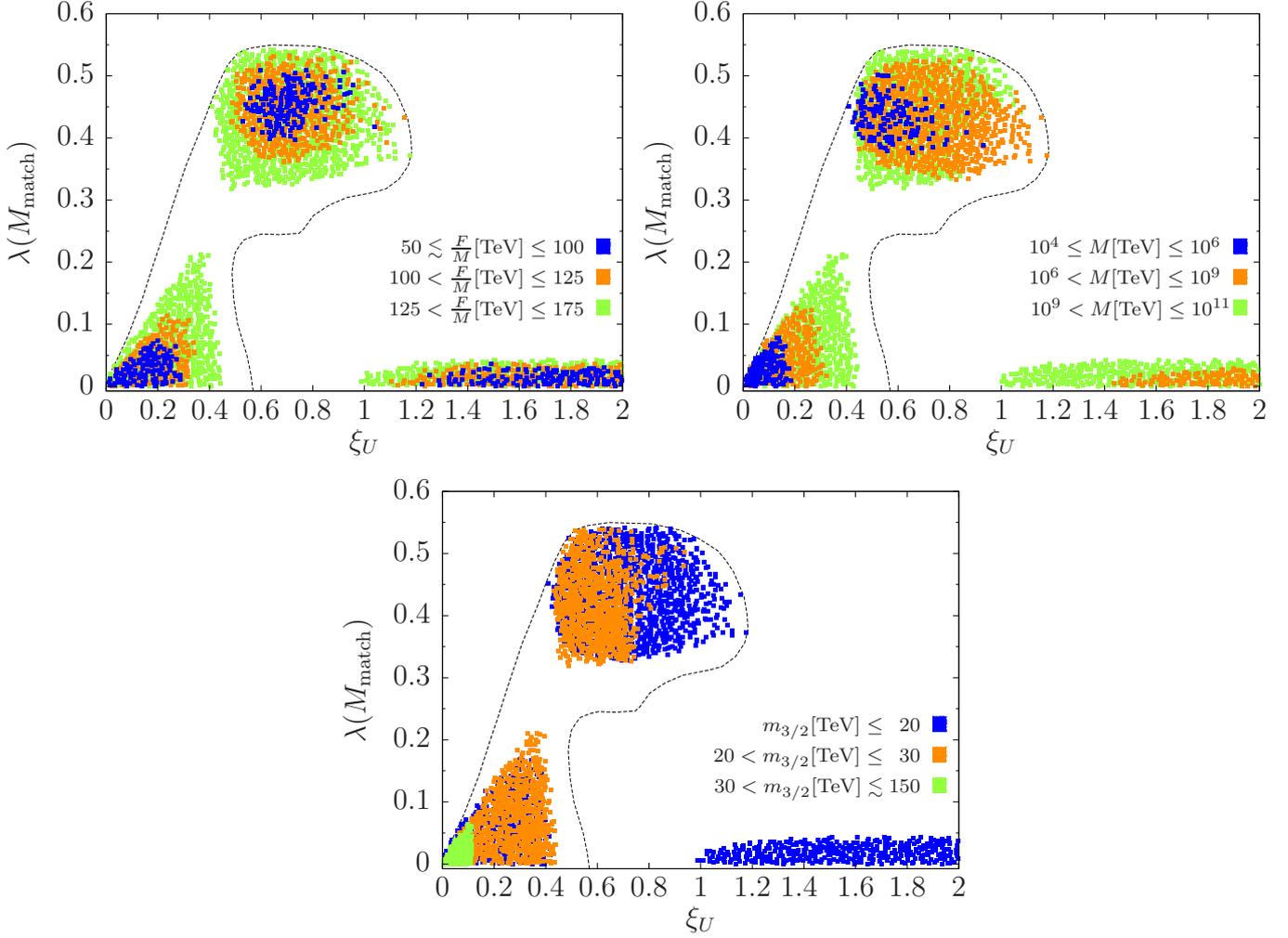}
\caption{(Upper-Left) Allowed regions in the $\lambda(M_{\mt{match}})$-$\xi_U$ plane for different ranges of values of the effective gauge-mediation SUSY breaking scale $F/M$.  (Upper-Right) The same for different ranges of values of the messenger scale $M$. (Lower) The same for different ranges of values of the gravitino mass $m_{3/2}$. Note that all the lower-left region is accessible for values of $m_{3/2}\leq 30 \units{TeV}$ . In all cases the dashed line bounds the region consistent with all collider bounds, but allowing also the possibility of a stau LSP.}
\label{fig_LXiU}
\end{figure}

The allowed regions in the $\lambda(M_\mt{match})$-$\xi_U$ plane are shown in Fig.~\!\ref{fig_LXiU}. We distinguish the points corresponding to different ranges of values for $F/M$ (upper-left panel), $M$ (upper-right panel), and $m_{3/2}$ (lower panel). As pointed out above, the most noticeable thing at first glance is that not too large values of the gravitino mass seem to be accepted in the scan. Thus, the shape and size of the allowed regions are very similar to those in \cite{Delgado:2007rz}, where AMSB was not included. At any rate, despite only a small deflection from anomaly mediation seems to be allowed, this suffices to change the phenomenological implications of the model. Note that, small values of $m_{3/2}$ diminish the potential of this model to avoid the fine tuning problem of \cite{Delgado:2007rz} for low SUSY breaking and messenger mass scales.

The reason for $m_{3/2}$ being small is that, for large values, in general only points where the right-handed (RH) stau is the lightest supersymmetric particle (LSP) are found. This follows from the fact that
a sizable $m_{3/2}$ implies not only lighter sleptons, but also heavier Binos and Winos compared to the gauge-mediation scenario. Thus, even if for large $m_{3/2}$ the gauge-mediation contributions manage to lift the tachyonic slepton masses above current bounds, the lightest neutralino tends to be heavier than the lightest stau, which is then the LSP. Such a heavy charged particle being stable would be in trouble with cosmological constraints. Therefore, although such points satisfy all the particle collider bounds, they are rejected in the scan. If we restrict to the requirement of EWSB and that the spectrum passes all present collider limits, significantly larger regions and values of $m_{3/2}$ would be allowed. Such regions are delimited by the solid line in Fig.~\!\ref{fig_LXiU}. In that case we find values of the gravitino mass distributed along the entire scan range,
\be
\left.m_{3/2}\right|_{{\tilde{\tau}_R}~\mt{LSP}}\leq \left.\frac{F}{M}\right|_\mt{max}=175 \units{TeV}.
\ee
Actually, in this case we can even obtain regions with $F/M<m_{3/2}$, where RG evolution effects down to $M_\mt{match}$ help in lifting the negative slepton masses squared. It is important to emphasize that these would be valid regions provided we introduce small sources of $R$-parity violation allowing the RH sleptons to decay, e.g. $\frac 12 \lambda_{ijk} l_i l_j e_k^c$ superpotential interactions.

Even if we allow a stau LSP, relatively large values of the gravitino mass, and then large AMSB contributions, are constrained to the left region in Fig.~\!\ref{fig_LXiU}. While, the lower-right region with large values of $\xi_U$ is populated only for $m_{3/2}\lesssim20\units{TeV}$. Indeed, as stressed in \cite{Delgado:2007rz}, in that region the singlet soft mass squared $m_S^2$ tends to be positive, and generating a large singlet vev is only possible through a large value of $a_\kappa$. Both $F/M$ and $m_{3/2}$ contributions give a negative $a_\kappa$, and can provide large values for a large $\xi_U$ and sizable $\kappa$ (see Eq.~\!(\ref{aterms})). However, large values of $\xi_U$, $\kappa$ and $m_{3/2}$ also increase the positive singlet mass squared, suppressing the singlet vev. Thus, for large $\xi_U$ the only viable solution implies $m_{3/2}\ll F/M$. Likewise, for relatively large values of $\lambda$ the gravitino mass cannot get too close to $F/M$. (In particular, $m_{3/2}\lesssim 80 \units{TeV}$ for $\lambda\gtrsim  0.4$.) Again, for large $\lambda$ the anomaly contribution increases the positive contribution to $m_S^2$. In this case the $a$ terms cannot get large enough values to generate the adequate vev for $S$.

Finally, as stressed above, it is apparent that the $R$-parity conserving scenario, allowing only for a small anomaly contribution, does not help in increasing the size of the allowed regions for low values of $F/M$ and $M$, compared to \cite{Delgado:2007rz}. Some improvement is possible if we consider the extended regions with a stau LSP. At any rate, we find that, compared to \cite{deBlas:2011hs}, the allowed regions in this model are somewhat smaller.

We now move to describe the general features of the spectrum. We have checked that in most of the parameter space the heavy Higgs bosons are essentially decoupled, so the lightest CP-even Higgs boson, $H^1$, is SM-like and must be heavier than the LEP 2 bound of $114\units{GeV}$ \cite{Barate:2003sz}. The exception is the small region at the lower-left corner of Fig.~\!\ref{fig_LXiU}, corresponding to very small values of $\lambda$ and $\xi_U$. In this case the singlet-like particles can be very light and the lightest neutral Higgses can have a large singlet component, escaping the LEP 2 bounds. On the other hand, $m_{H^1}\lesssim 123 \units{GeV}$. For $\tan{\beta}$ we find
\be
1.5\lesssim\tan{\beta}\lesssim 50,
\ee
although large values, $\tan{\beta}\gtrsim 10$, only occur in small regions (near the origin and the lower-right corner in Fig.~\!\ref{fig_LXiU}).

\begin{figure}[t]
% arXiv plots
\input{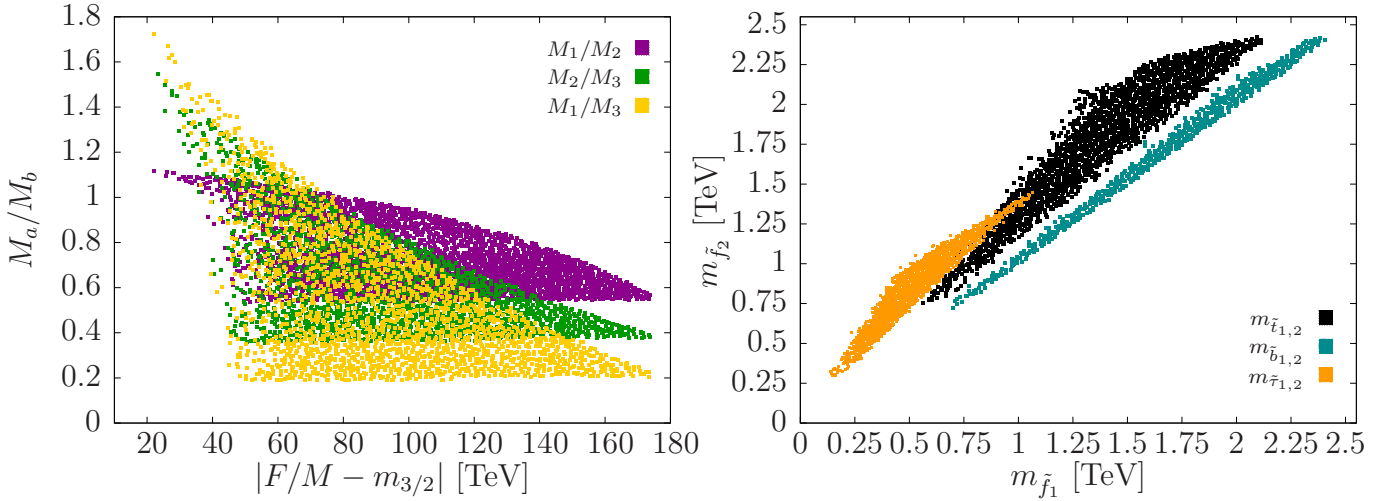}
\caption{(Left) Gaugino mass ratios as a funtion of the difference between the gauge and anomaly mediation scales: $|F/M-m_{3/2}|$. (Right) Third family sfermion masses. }
\label{fig_Spectrum}
\end{figure}

As in \cite{Delgado:2007rz,deBlas:2011hs} the singlet-messenger interactions do not alter significantly the spectrum of gaugino and sfermion masses. These can be, however, noticeably different than in the gauge-mediation scenario because of the anomaly contributions. As already noted above, since each mechanism has a different preference for the lightest gaugino, both Bino and Wino turn to be heavier than in the corresponding single scenarios. On the other hand, because of the negative sign of the $\beta$ function for $g_3$, the anomaly contribution lowers the gluino mass compared to gauge mediation. For large values of $m_{3/2}$, not too far from $F/M$, the gluino can be actually the lightest gaugino. This is illustrated in Fig.~\!\ref{fig_Spectrum}, left panel, where we show the values that the different gaugino mass ratios can take depending on the difference $|F/M-m_{3/2}|$. The plot corresponds to the $R$-parity conserving case, where in general only the splittings are affected. Still, this leads to a more compressed spectrum which might be challenging at the LHC.  Likewise, the ordering in the sfermion masses is essentially gauge mediation like, with anomaly-deflected splittings. The third family of sfermion masses are shown in Fig.~\!\ref{fig_Spectrum}, right panel. As can be observed, for instance, the lightest stop mass can be significantly below the TeV, as opposed to \cite{Delgado:2007rz}. 

Of course, another difference with respect to gauge mediation is that now the gravitino is heavier and it is no longer the LSP in general. This r{\^o}le 
corresponds to the lightest neutralino or, if $m_{3/2}$ is large enough, to the lightest sfermion: the RH stau.
The preference for $F/M > m_{3/2}$ also explains that in this model the lightest neutralino can be mostly Bino but not mostly Wino. Note also that, being both electroweak gauginos heavier than in each single scenario, a large Higgsino component can be present without requiring a too small $\mu^{\mt{Eff}}$ ($ {\cal O}\left(1\units{TeV}\right)$). Thus, the lightest neutralino is in general an admixture of Bino, Wino and Higgsinos. Sizable components of the last two, however, are restricted to the $R$-parity violating case. Finally, near the origin in the lower-left region in Fig.~\!\ref{fig_LXiU}, where the singlino can be very light, the lightest neutralino can be mostly singlino.

%-------------------------------- DOCUMENT: CONCLUSIONS----------------------------------%

\section{Conclusions}

Following previous works \cite{Delgado:2007rz,deBlas:2011hs}, in this paper we have studied a NMSSM-like scenario where supersymmetry is broken by soft terms induced by gauge mediation and extra singlet-messenger superpotential interactions. These are further deflected by contributions from anomaly mediation. While the gauge-mediation contributions lift the AMSB tachyonic sfermion masses, the singlet-messenger interactions allow for a realistic EWSB. The resulting scenario is free of the $\mu$-$b_\mu$ problem.

We have explored the parameter space of the model and found large regions where EWSB and a phenomenologically realistic spectrum are possible. For the model to work, the GMSB scale is in general required to be larger than the AMSB one, the gravitino mass. If we consider an $R$-parity conserving scenario the gravitino mass is actually restricted to take relatively small values, $m_{3/2}\lesssim 30 \units{TeV}$, due to cosmological constraints. Otherwise, for large values of $m_{3/2}$ we would end, in general, with a model with an stable charged LSP. 
Therefore, being $m_{3/2}$ small, the parameter space of this model is not much different than the corresponding gauge-mediation scenario. The same applies for the spectrum, which due to the anomaly contributions would be characterized by lighter sleptons, heavier squarks and a more compressed gaugino sector compared to gauge mediation. 

Small $R$-parity violating terms would help to enhance the differences between AGMSB and GMSB. Indeed, they allow to sidestep the cosmological problems, and thus to increase significantly the range of natural values for $m_{3/2}$. In particular, the size of the effects can modify the gauge-mediation gaugino mass order, which is not easy to attain in the $R$-parity conserving case. Including such terms also results in a larger size of the allowed regions in the parameter space, as shown in Fig.~\!\ref{fig_LXiU}. This would ameliorate the extra fine tuning problem found in the gauge-mediation case for low values of $F/M$ and $M$. At any rate, a large deflection from AMSB effects is possible only for not too large values of $\lambda$ and $\xi_U$.

%------------------------------------ ACKNOWLEDGEMENTS ---------------------------------------%
\section*{Acknowledgements}

This work has been supported in part by the U.S. National Science Foundation under Grant PHY-0905283-ARRA.

%------------------------------------------- REFERENCES -------------------------------------------%

%-------------------------------------------------- END --------------------------------------------------%


\begin{thebibliography}{99}

% Gravity Mediation

%\cite{Chamseddine:1982jx}
\bibitem{Chamseddine:1982jx}
  A.~H.~Chamseddine, R.~L.~Arnowitt and P.~Nath,
  %``Locally Supersymmetric Grand Unification,''
  Phys.\ Rev.\ Lett.\  {\bf 49}, 970 (1982);
  %%CITATION = PRLTA,49,970;%%
%
%\cite{Barbieri:1982eh}
%\bibitem{Barbieri:1982eh}
  R.~Barbieri, S.~Ferrara and C.~A.~Savoy,
  %``Gauge Models with Spontaneously Broken Local Supersymmetry,''
  Phys.\ Lett.\  B {\bf 119}, 343 (1982);
  %%CITATION = PHLTA,B119,343;%%
%
%\cite{Ibanez:1982ee}
%\bibitem{Ibanez:1982ee}
  L.~E.~Ibanez,
  %``Locally Supersymmetric SU(5) Grand Unification,''
  Phys.\ Lett.\  B {\bf 118}, 73 (1982);
  %%CITATION = PHLTA,B118,73;%%
%
%\cite{Hall:1983iz}
%\bibitem{Hall:1983iz}
  L.~J.~Hall, J.~D.~Lykken and S.~Weinberg,
  %``Supergravity as the Messenger of Supersymmetry Breaking,''
  Phys.\ Rev.\  D {\bf 27}, 2359 (1983);
  %%CITATION = PHRVA,D27,2359;%%
%
%\cite{Ohta:1982wn}
%\bibitem{Ohta:1982wn}
  N.~Ohta,
  %``GRAND UNIFIED THEORIES BASED ON LOCAL SUPERSYMMETRY,''
  Prog.\ Theor.\ Phys.\  {\bf 70}, 542 (1983);
  %%CITATION = PTPKA,70,542;%%
%
%\cite{Ellis:1982wr}
%\bibitem{Ellis:1982wr}
  J.~R.~Ellis, D.~V.~Nanopoulos and K.~Tamvakis,
  %``Grand Unification in Simple Supergravity,''
  Phys.\ Lett.\  B {\bf 121}, 123 (1983);
  %%CITATION = PHLTA,B121,123;%%
%
%\cite{AlvarezGaume:1983gj}
%\bibitem{AlvarezGaume:1983gj}
  L.~Alvarez-Gaume, J.~Polchinski and M.~B.~Wise,
  %``Minimal Low-Energy Supergravity,''
  Nucl.\ Phys.\  B {\bf 221}, 495 (1983).
  %%CITATION = NUPHA,B221,495;%%

% Gauge Mediation

%\cite{Dine:1981gu}
\bibitem{Dine:1981gu}
  M.~Dine and W.~Fischler,
  %``A Phenomenological Model Of Particle Physics Based On Supersymmetry,''
  Phys.\ Lett.\  B {\bf 110}, 227 (1982);
  %%CITATION = PHLTA,B110,227;%%
%
%\cite{Nappi:1982hm}
%\bibitem{Nappi:1982hm}
  C.~R.~Nappi and B.~A.~Ovrut,
  %``Supersymmetric Extension Of The SU(3) X SU(2) X U(1) Model,''
  Phys.\ Lett.\  B {\bf 113}, 175 (1982);
  %%CITATION = PHLTA,B113,175;%%
%  
%\cite{AlvarezGaume:1981wy}
%\bibitem{AlvarezGaume:1981wy}
  L.~Alvarez-Gaume, M.~Claudson and M.~B.~Wise,
  %``Low-Energy Supersymmetry,''
  Nucl.\ Phys.\  B {\bf 207}, 96 (1982);
  %%CITATION = NUPHA,B207,96;%%
%  
%\cite{Dine:1993yw}
%\bibitem{Dine:1993yw}
  M.~Dine and A.~E.~Nelson,
  %``Dynamical supersymmetry breaking at low-energies,''
  Phys.\ Rev.\  D {\bf 48}, 1277 (1993)
  [arXiv:hep-ph/9303230];
  %%CITATION = PHRVA,D48,1277;%%
%  
%\cite{Dine:1994vc}
%\bibitem{Dine:1994vc}
  M.~Dine, A.~E.~Nelson and Y.~Shirman,
  %``Low-Energy Dynamical Supersymmetry Breaking Simplified,''
  Phys.\ Rev.\  D {\bf 51}, 1362 (1995)
  [arXiv:hep-ph/9408384];
  %%CITATION = PHRVA,D51,1362;%%
%  
%\cite{Dine:1995ag}
%\bibitem{Dine:1995ag}
  M.~Dine, A.~E.~Nelson, Y.~Nir and Y.~Shirman,
  %``New tools for low-energy dynamical supersymmetry breaking,''
  Phys.\ Rev.\  D {\bf 53}, 2658 (1996)
  [arXiv:hep-ph/9507378];
  %%CITATION = PHRVA,D53,2658;%%
%  
%\cite{Giudice:1998bp}
%\bibitem{Giudice:1998bp}
  G.~F.~Giudice and R.~Rattazzi,
  %``Theories with gauge-mediated supersymmetry breaking,''
  Phys.\ Rept.\  {\bf 322}, 419 (1999)
  [arXiv:hep-ph/9801271].
  %%CITATION = PRPLC,322,419;%%
  
% Anomaly mediation

%\cite{Randall:1998uk}
\bibitem{Randall:1998uk}
  L.~Randall and R.~Sundrum,
  %``Out of this world supersymmetry breaking,''
  Nucl.\ Phys.\  B {\bf 557}, 79 (1999)
  [arXiv:hep-th/9810155].
  %%CITATION = NUPHA,B557,79;%%

%\cite{Giudice:1998xp}
\bibitem{Giudice:1998xp}
  G.~F.~Giudice, M.~A.~Luty, H.~Murayama and R.~Rattazzi,
  %``Gaugino mass without singlets,''
  JHEP {\bf 9812}, 027 (1998)
  [arXiv:hep-ph/9810442].
  %%CITATION = JHEPA,9812,027;%%

% Mu problem in gravity mediation
  
%\cite{Giudice:1988yz}
\bibitem{Giudice:1988yz}
  G.~F.~Giudice and A.~Masiero,
  %``A Natural Solution to the mu Problem in Supergravity Theories,''
  Phys.\ Lett.\  B {\bf 206}, 480 (1988).
  %%CITATION = PHLTA,B206,480;%%
  
% Superconformal SUGRA 
  
%\cite{Cremmer:1978hn}
\bibitem{Cremmer:1978hn}
  E.~Cremmer, B.~Julia, J.~Scherk, S.~Ferrara, L.~Girardello and P.~van Nieuwenhuizen,
  %``Spontaneous Symmetry Breaking and Higgs Effect in Supergravity Without
  %Cosmological Constant,''
  Nucl.\ Phys.\  B {\bf 147}, 105 (1979);
  %%CITATION = NUPHA,B147,105;%%
%
%\cite{Cremmer:1982en}
%\bibitem{Cremmer:1982en}
  E.~Cremmer, S.~Ferrara, L.~Girardello and A.~Van Proeyen,
  %``Yang-Mills Theories with Local Supersymmetry: Lagrangian, Transformation
  %Laws and SuperHiggs Effect,''
  Nucl.\ Phys.\  B {\bf 212}, 413 (1983).
  %%CITATION = NUPHA,B212,413;%%
  
% Combining Anomaly with (to solve tachyonic slepton problems and others...)

%Anomaly/Gauge mediation

%\cite{Pomarol:1999ie}
\bibitem{Pomarol:1999ie}
  A.~Pomarol and R.~Rattazzi,
  %``Sparticle masses from the superconformal anomaly,''
  JHEP {\bf 9905}, 013 (1999)
  [arXiv:hep-ph/9903448];
  %%CITATION = JHEPA,9905,013;%%
%  
%\cite{Hsieh:2006ig}
%\bibitem{Hsieh:2006ig}
  K.~Hsieh and M.~A.~Luty,
  %``Mixed gauge and anomaly mediation from new physics at 10-TeV,''
  JHEP {\bf 0706}, 062 (2007)
  [arXiv:hep-ph/0604256];
  %%CITATION = JHEPA,0706,062;%%
%  
%\cite{Cai:2010tj}
%\bibitem{Cai:2010tj}
  Y.~Cai and M.~A.~Luty,
  %``Minimal Gaugomaly Mediation,''
  JHEP {\bf 1012}, 037 (2010)
  [arXiv:1008.2024 [hep-ph]].
  %%CITATION = JHEPA,1012,037;%%

% Others

%\cite{AnomalyPlus}
\bibitem{AnomalyPlus}
%
%\cite{Chacko:1999am}
%\bibitem{Chacko:1999am}
  Z.~Chacko, M.~A.~Luty, I.~Maksymyk and E.~Ponton,
  %``Realistic anomaly mediated supersymmetry breaking,''
  JHEP {\bf 0004}, 001 (2000)
  [arXiv:hep-ph/9905390];
  %%CITATION = JHEPA,0004,001;%%
%
%\cite{Jack:2000cd}
%\bibitem{Jack:2000cd}
  I.~Jack and D.~R.~T.~Jones,
  %``Fayet-Iliopoulos D terms and anomaly mediated supersymmetry breaking,''
  Phys.\ Lett.\  B {\bf 482}, 167 (2000)
  [arXiv:hep-ph/0003081];
  %%CITATION = PHLTA,B482,167;%%
%
%\cite{Kaplan:2000jz}
%\bibitem{Kaplan:2000jz}
  D.~E.~Kaplan and G.~D.~Kribs,
  %``Gaugino-assisted anomaly mediation,''
  JHEP {\bf 0009}, 048 (2000)
  [arXiv:hep-ph/0009195];
  %%CITATION = JHEPA,0009,048;%%
%  
%\cite{Murakami:2003pb}
%\bibitem{Murakami:2003pb}
  B.~Murakami and J.~D.~Wells,
  %``Abelian D terms and the superpartner spectrum of anomaly mediated
  %supersymmetry breaking,''
  Phys.\ Rev.\  D {\bf 68}, 035006 (2003)
  [arXiv:hep-ph/0302209];
  %%CITATION = PHRVA,D68,035006;%%
%  
%\cite{Sundrum:2004un}
%\bibitem{Sundrum:2004un}
  R.~Sundrum,
  %``'Gaugomaly' mediated SUSY breaking and conformal sequestering,''
  Phys.\ Rev.\  D {\bf 71}, 085003 (2005)
  [arXiv:hep-th/0406012];
  %%CITATION = PHRVA,D71,085003;%%
%
%\cite{deBlas:2009vx}
%\bibitem{deBlas:2009vx}
  J.~de Blas, P.~Langacker, G.~Paz and L.~T.~Wang,
  %``Combining Anomaly and Z' Mediation of Supersymmetry Breaking,''
  JHEP {\bf 1001}, 037 (2010)
  [arXiv:0911.1996 [hep-ph]].
  %%CITATION = JHEPA,1001,037;%% 

% NMSSM and GM
  
%\cite{deGouvea:1997cx}
\bibitem{deGouvea:1997cx}
  A.~de Gouvea, A.~Friedland and H.~Murayama,
  %``Next-to-minimal supersymmetric standard model with the gauge mediation  of
  %supersymmetry breaking,''
  Phys.\ Rev.\  D {\bf 57}, 5676 (1998)
  [arXiv:hep-ph/9711264].
  %%CITATION = PHRVA,D57,5676;%%
  
% NMSSM and AM  

%\cite{Kitano:2004zd}
\bibitem{Kitano:2004zd}
  R.~Kitano, G.~D.~Kribs and H.~Murayama,
  %``Electroweak symmetry breaking via UV insensitive anomaly mediation,''
  Phys.\ Rev.\  D {\bf 70}, 035001 (2004)
  [arXiv:hep-ph/0402215];
  %%CITATION = PHRVA,D70,035001;%%
  
% Model with Singlet messenger interactions

%\cite{Giudice:1997ni}
\bibitem{Giudice:1997ni}
  G.~F.~Giudice and R.~Rattazzi,
  %``Extracting Supersymmetry-Breaking Effects from Wave-Function
  %Renormalization,''
  Nucl.\ Phys.\  B {\bf 511}, 25 (1998)
  [arXiv:hep-ph/9706540].
  %%CITATION = NUPHA,B511,25;%%

%\cite{Delgado:2007rz}
\bibitem{Delgado:2007rz}
  A.~Delgado, G.~F.~Giudice and P.~Slavich,
  %``Dynamical mu Term in Gauge Mediation,''
  Phys.\ Lett.\  B {\bf 653}, 424 (2007)
  [arXiv:0706.3873 [hep-ph]].
  %%CITATION = PHLTA,B653,424;%%
  
% Singlet deflection

%\cite{deBlas:2011hs}
\bibitem{deBlas:2011hs}
  J.~de Blas and A.~Delgado,
  %``Exploring singlet deflection of gauge mediation,''
  Phys.\ Rev.\  D {\bf 83}, 115011 (2011)
  [arXiv:1103.3280 [hep-ph]].
  %%CITATION = PHRVA,D83,115011;%%

% Suppression of gravity mediation
  
%\cite{Luty:1999cz}
\bibitem{Luty:1999cz}
  M.~A.~Luty and R.~Sundrum,
  %``Radius stabilization and anomaly mediated supersymmetry breaking,''
  Phys.\ Rev.\  D {\bf 62}, 035008 (2000)
  [arXiv:hep-th/9910202].
  %%CITATION = PHRVA,D62,035008;%%

% NMSSM RGEs  
  
%\cite{Ellwanger:2009dp}
\bibitem{Ellwanger:2009dp}
  U.~Ellwanger, C.~Hugonie and A.~M.~Teixeira,
  %``The Next-to-Minimal Supersymmetric Standard Model,''
  Phys.\ Rept.\  {\bf 496}, 1 (2010)
  [arXiv:0910.1785 [hep-ph]].
  %%CITATION = PRPLC,496,1;%%
  
% NMHDECAY
  
%\cite{Ellwanger:2004xm}
\bibitem{Ellwanger:2004xm}
  U.~Ellwanger, J.~F.~Gunion and C.~Hugonie,
  %``NMHDECAY: A Fortran code for the Higgs masses, couplings and decay widths
  %in the NMSSM,''
  JHEP {\bf 0502}, 066 (2005)
  [arXiv:hep-ph/0406215]; 
  %%CITATION = JHEPA,0502,066;%%
%\cite{Ellwanger:2005dv}
%\bibitem{Ellwanger:2005dv}
  U.~Ellwanger and C.~Hugonie,
  %``NMHDECAY 2.0: An Updated program for sparticle masses, Higgs masses,
  %couplings and decay widths in the NMSSM,''
  Comput.\ Phys.\ Commun.\  {\bf 175}, 290 (2006)
  [arXiv:hep-ph/0508022].
  %%CITATION = CPHCB,175,290;%%
  
% LEP 2 bound on MH
  
%\cite{Barate:2003sz}
\bibitem{Barate:2003sz}
  R.~Barate {\it et al.}  [LEP Working Group for Higgs boson searches and
                  ALEPH Collaboration and  and],
  %``Search for the standard model Higgs boson at LEP,''
  Phys.\ Lett.\  B {\bf 565}, 61 (2003)
  [arXiv:hep-ex/0306033].
  %%CITATION = PHLTA,B565,61;%% 

\end{thebibliography}
\end{document}